\pdfoutput=1
\documentclass[prl,twocolumn,showpacs,amsmath,amssymb,superscriptaddress,floatfix]{revtex4}

\usepackage{graphicx}
\usepackage{dcolumn}
\usepackage{bm}
\usepackage{epsfig}

\begin{document}

\title{Quantum-interference-controlled three-terminal molecular transistors based on a single ring-shaped-molecule connected to graphene nanoribbon electrodes}

\author{Kamal K. Saha}
\affiliation{Department of Physics and Astronomy, University of Delaware, Newark, DE 19716-2570, USA}
\author{Branislav K. Nikoli\' c}
\email{bnikolic@udel.edu}
\affiliation{Department of Physics and Astronomy, University of Delaware, Newark, DE 19716-2570, USA}
\affiliation{Institute for Theoretical Physics, University of Regensburg, 93040 Regensburg, Germany}
\author{Vincent Meunier}
\affiliation{Department of Physics, Applied Physics, and Astronomy, Rensselaer Polytechnic Institute, 110 Eighth Street, Troy, New York 12180-3590, USA}
\author{Wenchang Lu}
\affiliation{Center for High Performance Simulation and Department of Physics, North Carolina State University,
Raleigh, North Carolina 27695-7518, USA}
\author{J. Bernholc}
\affiliation{Center for High Performance Simulation and Department of Physics, North Carolina State University, Raleigh, North Carolina 27695-7518, USA}

\begin{abstract}
We study all-carbon-hydrogen molecular transistors where zigzag graphene nanoribbons play the role of three metallic electrodes connected to a ring-shaped 18-annulene molecule. Using the nonequilibrium Green function formalism combined with density functional theory, recently extended to multiterminal devices, we show that the proposed nanostructures exhibit exponentially small transmission when the source and drain electrodes are attached in a configuration that ensures destructive interference of electron paths around the ring. The third electrode, functioning either as an attached infinite-impedance voltage probe or as  an ``air-bridge'' top gate covering half of molecular ring, introduces dephasing that brings the transistor into the `on' state with its transmission in the latter case approaching the maximum limit for a single conducting channel device. The current through the latter device can also be controlled in the far-from-equilibrium regime by applying a gate voltage.
\end{abstract}

\pacs{85.35.Ds, 85.65.+h, 73.63.Rt, 72.80.Vp}
\maketitle

{\em Introduction.}---The recent fabrication~\cite{Wang2008a,Cai2010} of graphene nanoribbons (GNRs) with ultrasmooth edges has opened unforeseen avenues for nanoelectronics~\cite{Schwierz2010} by providing a novel type of semiconducting channel (GNRs with armchair edges of sub-10-nm width and band gaps $\gtrsim 0.4$ eV) for field-effect transistors (FETs)~\cite{Wang2008a} or low-dissipation interconnects~\cite{Areshkin2007a} (metallic GNRs with zigzag edges). Furthermore,   metallic GNRs have the capability to resolve one of the key challenges~\cite{Kiguchi2008} for molecular electronics~\cite{Cuniberti2005}---{\em a well-defined molecule-electrode contact with high transparency, strong directionality and reproducibility}.

This is due to the fact that strong molecule-GNR \mbox{$\pi$-$\pi$} coupling makes possible  formation of a continuous \mbox{$\pi$-bonded} network across GNR and orbitals of conjugated organic molecules~\cite{Ke2007}. In addition, unusual electronic structure~\cite{Areshkin2007a} of zigzag GNRs (ZGNRs) and the possibility to control it via edge doping~\cite{Cervantes-Codi2008} offer prospect for device designs that are outside of the scope of traditional approaches in molecular electronics where single molecules are in contact with metals~\cite{Kiguchi2008,Song2009}, such as gold or platinum which suffer from the lack of directional covalent bonding thereby leading to poor reproducibility of most metal-molecule-metal junctions~\cite{Stokbro2003}, or carbon nanotubes~\cite{Guo2006} where edges are not present and which do not have planar structure appropriate for aligning and patterning through lithographic methods suitable for high-volume production.

Unlike carbon nanotubes (CNTs), which were employed experimentally~\cite{Guo2006} as electrodes of molecular devices and whose contact to conjugated molecules  has been investigated via first-principles quantum transport methods~\cite{Ke2007}, surprisingly little is known about linear response or far-from-equilibrium transport properties of multiterminal molecule-GNR heterojunctions. Although pristine ZGNRs have a band gap at low temperature due to edge magnetic ordering~\cite{Yazyev2008,Areshkin2009}, this gap is easily destroyed at room temperature~\cite{Yazyev2008} or by defects and impurities along the edge~\cite{Areshkin2009,Jiang2008}.

To obtain transistor effect,  current through the molecular junction has to be controlled by the third electrode acting either as an electrostatically coupled gate~\cite{Schwierz2010} in traditional FET designs~\cite{Song2009} or as being attached to the molecule in quantum-interference-controlled transistor (QICT) concepts~\cite{Cardamone2006}. In the recently proposed QICT, source and drain electrodes are connected to a ring-shaped molecule in a configuration that ensures destructive interference~\cite{Nazarov2009} of electron paths around the ring and, therefore, `off' state of QICT with perfect zero leakage current. The third electrode, acting as the so-called B\"uttiker voltage probe which does not draw any current~\cite{Nazarov2009}, is then attached to the molecule to introduce dephasing and thereby `on' state of QICT.

The QICT also offers a playground for basic research on quantum interference effects~\cite{Nazarov2009} in molecular electronics which, unlike in conventional two-dimensional electron gases or graphene, can manifest itself even at room temperature since molecular vibrations with dephasing effect (i.e., those that change the length difference between paths around the ring) can be suppressed~\cite{Cardamone2006} below 500 K. Such interference effects in transport through aromatic molecules have attracted considerable attention in the very recent experimental and theoretical studies~\cite{Rincon2009}. Furthermore, potential applications of QICT involve operation with greatly reduced heat dissipation because the current flow is not blocked by an energy barrier as in traditional FETs where it must be raised and lowered with each switching cycle in either silicon-based or carbon-based FETs~\cite{Schwierz2010}.

\begin{figure}
\includegraphics[scale=0.33,angle=0]{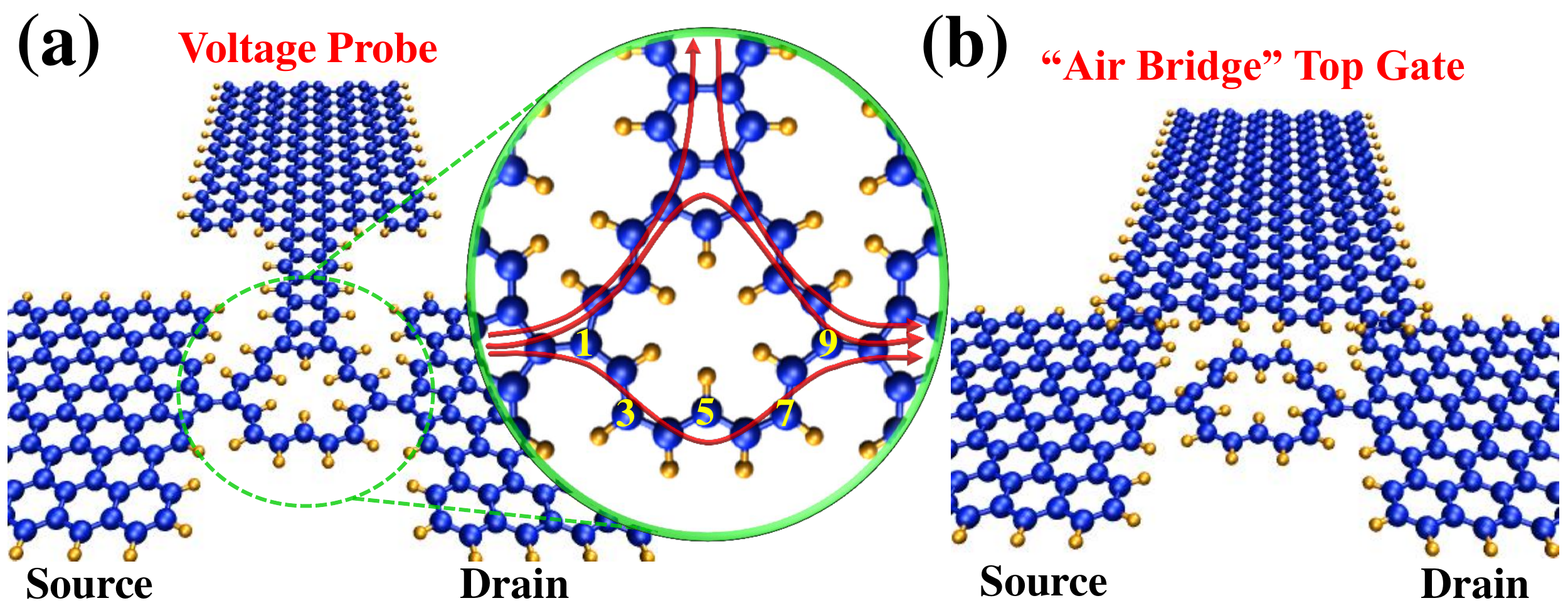}
\caption{(Color online) Schematic view of the proposed ZGNR$|$18-annulene$|$ZGNR three-terminal heterojuctions. The contact between the source and drain 8-ZGNR metallic electrodes and a ring-shaped 18-annulene molecule is made via 5-membered rings of carbon atoms (dark blue), while the electrodes are attached in a destructive configuration (1,9) for quantum interference. The source-drain current is controlled by a third electrode which is coupled to the device either through bonding via  5-membered ring (a) or as an ``air-bridge'' top gate covering upper half of the molecular ring (b). The hydrogen atoms (light yellow) are included to passivate the edge carbon atoms. The inset shows carbon atom numbering within the molecular ring, as well as three possible Feynman paths of a single electron entering the ring.}
\label{fig:setup}
\end{figure}

Understanding of realistic junctions of ZGNR and molecules requires quantum transport methods combined with the {\em first-principles} input about atomistic and electronic structure to capture {\em charge transfer} in equilibrium (which is indispensable to obtain correct linear response conductance of, e.g., carbon-hydrogen systems~\cite{Areshkin2010}) or {\em charge redistribution} (whose knowledge ensures the gauge invariance of the current-voltage characteristics) in the far-from-equilibrium regime driven by finite bias voltage. The state-of-the-art approach that can capture all of these effects, as long as the coupling between the molecule and the electrodes is strong enough to diminish Coulomb blockade effects~\cite{Cuniberti2005}, is the nonequilibrium Green function formalism combined with density functional theory (NEGF-DFT)~\cite{Cuniberti2005,Areshkin2010}. However, due to the lack of NEGF-DFT algorithms for {\em multiterminal} devices, previous efforts to model QICT based on benzene-gold~\cite{Cardamone2006,Ke2008}, annulene-gold~\cite{Ke2008}, benzene-CNT~\cite{Ke2008} and annulene-CNT~\cite{Ke2008} junctions have employed either semi-empirical models~\cite{Cardamone2006} (with truncated basis set and the connection between the leads and the molecule treated phenomenologically, thereby ignoring possible structural relaxation or hybridization between molecular $\pi$ states and contact $\sigma$ states) or NEGF-DFT~\cite{Ke2008} but in the absence of third attached electrode.

\begin{figure}
\includegraphics[scale=0.34,angle=0]{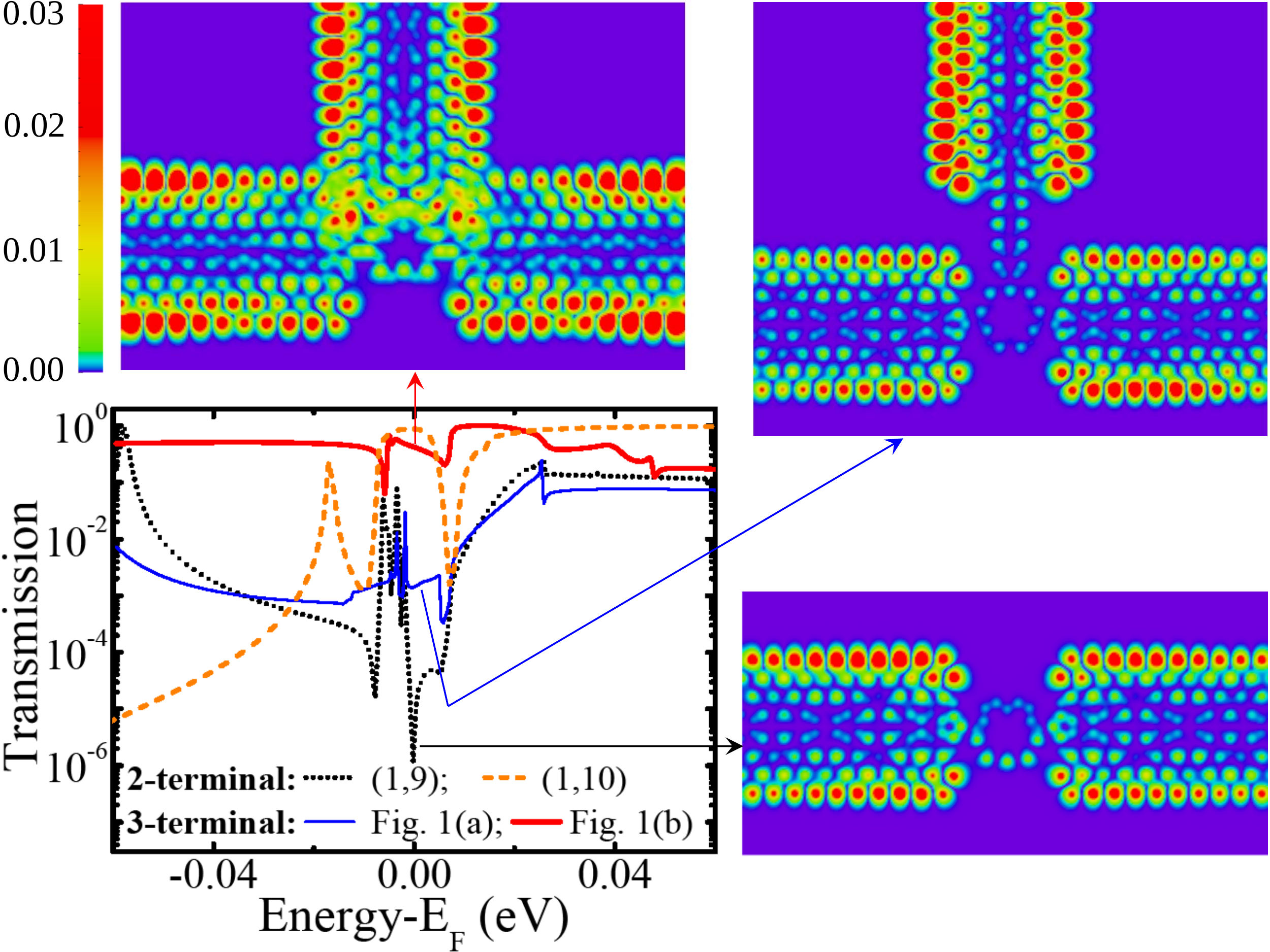}
\caption{(Color online) The effective source-drain transmission $\tilde{T}_{21}(E)$ for three-terminal QICTs shown in Fig.~\ref{fig:setup} and the corresponding LDOS at the Fermi energy. For comparison, we also plot transmission $T_{21}(E)$ for two-terminal devices in configurations (1,9) and (1,10) exhibiting  destructive and constructive quantum interference, respectively.}
\label{fig:transmission}
\end{figure}

\begin{figure}
\includegraphics[scale=0.33,angle=0]{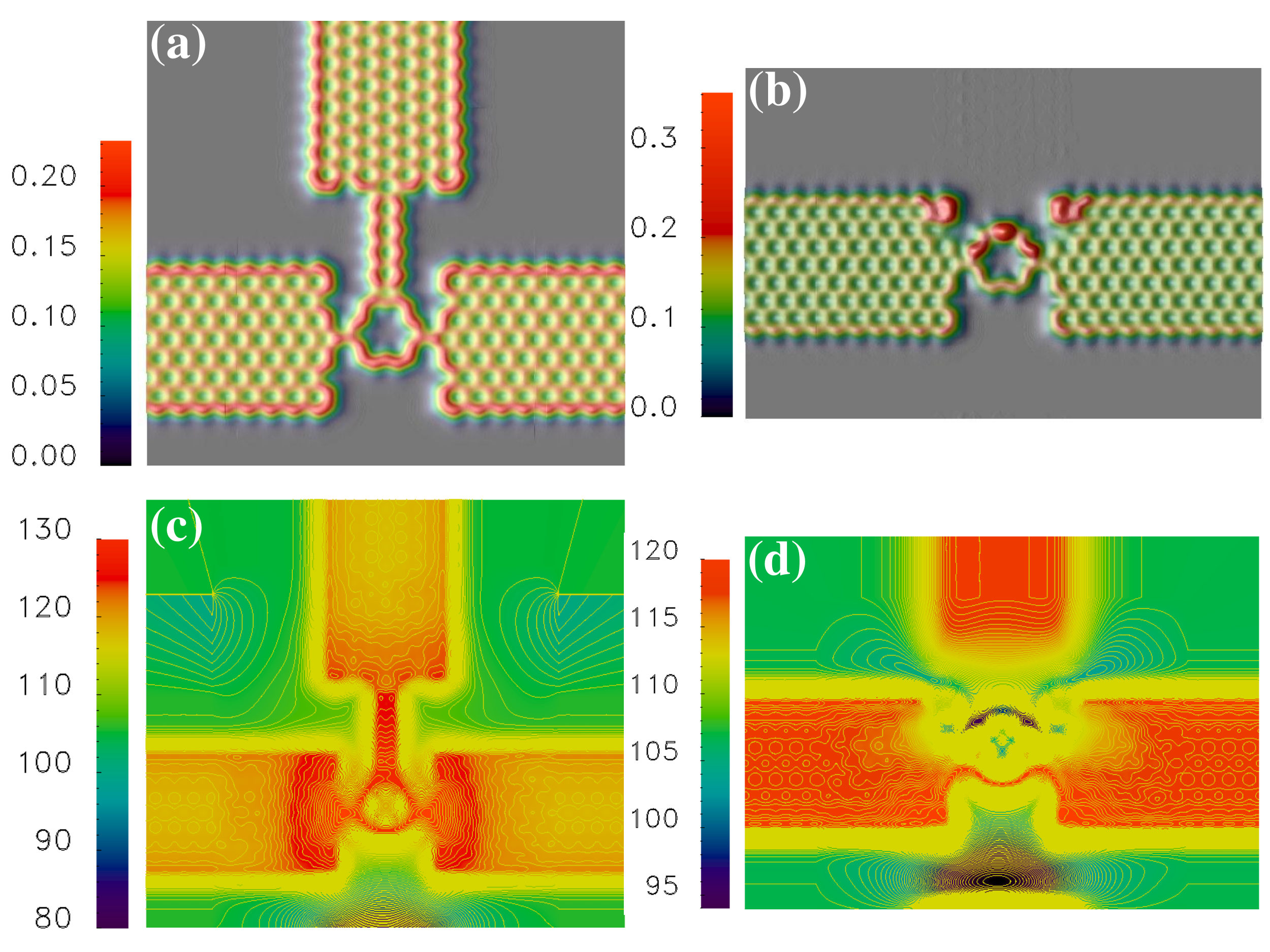}
\caption{(Color online) The self-consistent equilibrium charge density [(a) and (b)] and the Hartree potential [(c) and (d)] computed within the central region
of size  of size $62 \ \mbox{\AA} \ \times \ 50 \ \mbox{\AA}$ of the three-terminal ZGNR$|$18-annulene$|$ZGNR devices shown in Fig.~\ref{fig:setup}. The third electrode is either directly attached to the 18-annulene molecule [(a) and (c)], or  it acts as ``air-bridge'' top gate at a distance $3.2$ \mbox{\AA} away from the upper half of the molecular ring which it covers [(b) and (d)].
}
\label{fig:charge}
\end{figure}

\begin{figure}
\includegraphics[scale=0.25,angle=0]{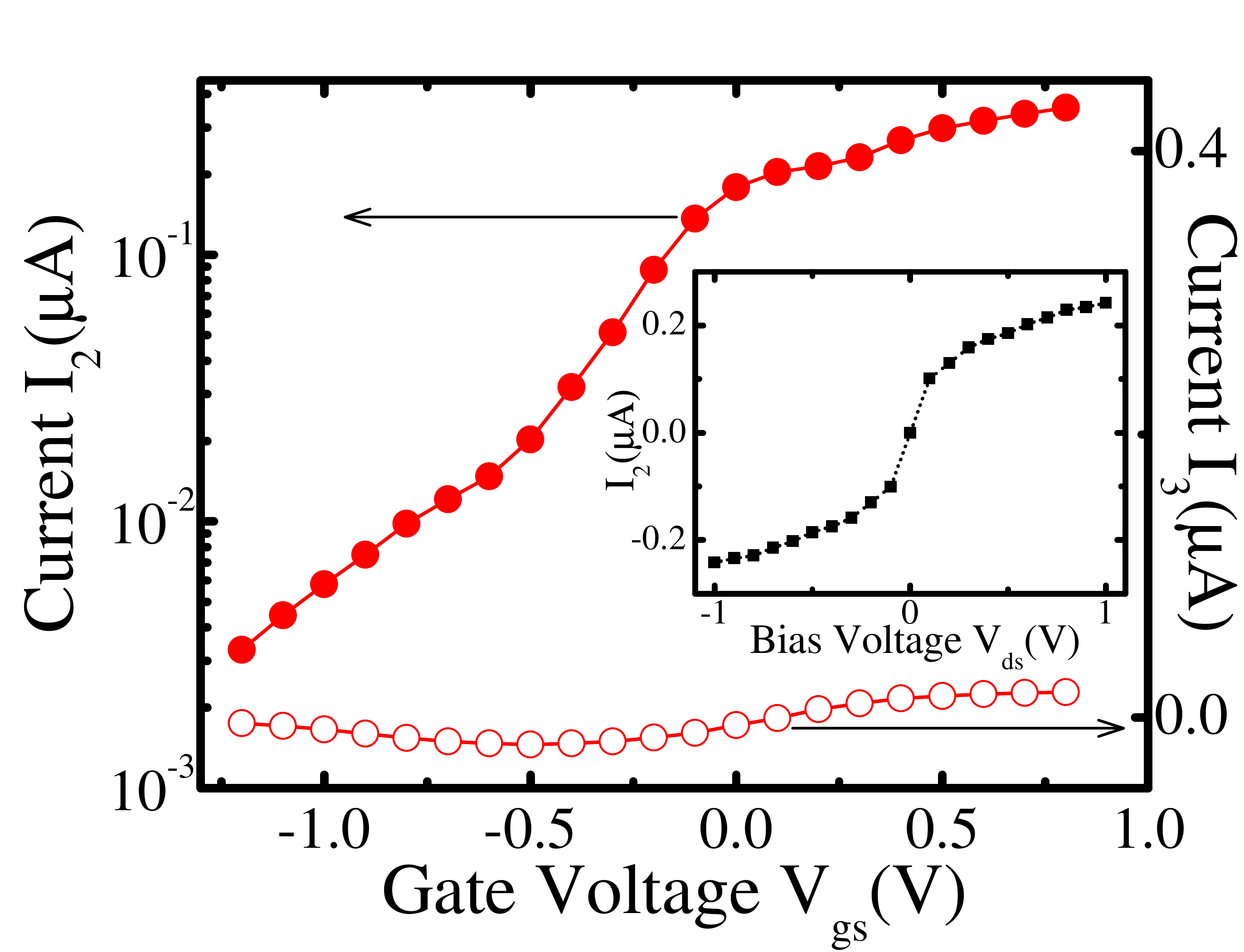}
\caption{(Color online) Transfer characteristics showing drain current $I_2$ versus the gate-source voltage $V_{gs}$ for the device in Fig.~\ref{fig:setup}(b) biased by the drain-source voltage \mbox{$V_{ds}=0.2$ V}. Right axis gives values of current in the third (top gate) electrode.  The inset plots $I_2$ vs. $V_{ds}$  for the two-terminal device.}
\label{fig:finitebias}
\end{figure}

In this Letter, we analyze linear response and nonequilibrium transport properties of \mbox{ZGNR$|$18-annulene$|$ZGNR junctions}, whose third electrode is made of ZGNR as well, by employing the very recently developed NEGF-DFT formalism for multiterminal nanostructures (MT-NEGF-DFT)~\cite{Saha2009a}. The third electrode introduces dephasing by being attached to the molecule to act as an infinite-impedance voltage probe~\cite{Nazarov2009}, as shown in Fig.~\ref{fig:setup}(a). Moreover, we find an even larger disruption of destructive quantum interference in a setup proposed in Fig.~\ref{fig:setup}(b) where the third electrode plays the role of a top gate covering the upper-half of annulene ring while being separated from the two terminal device underneath by an air gap. Our principal result is shown in Fig.~\ref{fig:transmission} where exponentially small  transmission $T_{21}(E_F) \simeq 10^{-6}$ between the source electrode 1 and the drain electrode 2 at the Fermi energy $E_F$ in the  junction with destructive geometry for connecting the two ZGNR electrodes increases by three orders of magnitude $\tilde{T}_{21}(E_F) \simeq 10^{-3}$  with the attachment of an infinite-impedance voltage probe. Moreover, the junction becomes highly transparent $\tilde{T}_{21}(E_F) \simeq 0.4$ due to induced states (Fig.~\ref{fig:transmission}) and charge transfer (Fig.~\ref{fig:charge}) underneath the ``air-bridge'' top gate. We also analyze in Fig.~\ref{fig:finitebias} the gate voltage modulation of source-drain current in the far-from-equilibrium transport regime of device in Fig.~\ref{fig:setup}(b).

{\em Device setup and MT-NEGF-DFT formalism.}---Although na\"ively one would expect that formation of continuous $\pi$-bonded network between carbon-based electrodes and conjugated molecules would ensure high contact transparency, early experiments~\cite{Guo2006} on CNT$|$molecule$|$CNT heterojunctions  have measured surprisingly small conductances for a variety of  sandwiched molecules suggesting poor contact transparency. The first-principles analysis of different setups reveals that this is due to significant twisting forces when molecule is connected to CNT via, e.g., 6-membered rings~\cite{Ke2007}. Therefore, to keep nearly parallel and in-plane configuration (hydrogen atoms of annulene slightly deviate from the molecular plane) of our ZGNR$|$18-annulene$|$ZGNR junction, we use a 5-membered ring~\cite{Ke2007} in Fig.~\ref{fig:setup} to chemically bond ZGNR to annulene. The atomic structure of the heterojunctions  in Fig.~\ref{fig:setup} is fully optimized by minimizing the atomic forces on individual atoms to be smaller than $0.05$ eV/\AA. This preserves the overall conjugation and leads to strong coupling and {\em high contact transparency}, as confirmed by Fig.~\ref{fig:transmission} where $T_{21}(E_F) \simeq 0.9$ for ZGNR electrodes attached to annulene atoms (1,10)  as an example of configuration with constructive quantum interference.

The high contact transparency also makes it possible to apply the NEGF-DFT framework, which would otherwise be rendered insufficient due to  electronic  correlations~\cite{Darau2009} emerging in the weak coupling regime that are beyond the mean-field DFT treatment~\cite{Cuniberti2005}. In the NEGF-DFT formalism, the Hamiltonian is not known in advance and has to be computed by finding converged charge redistribution via the self-consistent DFT loop for the density matrix ${\bm \rho} = \frac{1}{2 \pi i} \int dE\, {\bf G}^<(E)$ whose trace gives the charge density~\cite{Cuniberti2005,Areshkin2010}. The NEGF formalism for steady-state transport operates with two central quantities, retarded ${\bf G}(E)$ and lesser ${\bf G}^<(E)$ Green functions, which describe the density of available quantum states and how electrons occupy those states, respectively. The technical details of the construction of the nonequilibrium density matrix ${\bm \rho}$ for multiterminal devices are discussed in Ref.~\cite{Saha2009a}. In Fig.~\ref{fig:charge}, we show the equilibrium charge density and the Hartree potential (obtained by solving the Poisson equation in the scattering region with the boundary conditions that match the electrostatic potentials of all three electrodes) computed through self-consistent loop for two QICTs from Fig.~\ref{fig:setup}.

In the coherent transport regime (i.e., in the absence of electron-phonon or electron-electron dephasing processes), the NEGF post-processing of the result of the DFT loop expresses the current flowing into terminal $\alpha$ of the device as:
\begin{equation}\label{eq:current}
I_\alpha = \frac{2e}{h} \sum_{\beta=1}^{3} \int\limits_{-\infty}^{+\infty} dE\, T_{\beta\alpha}(E,V_\alpha,V_\beta) [f_\alpha(E)-f_\beta(E)].
\end{equation}
Here the transmission coefficients $T_{\beta\alpha}(E,V_\alpha,V_\beta) = {\rm Tr} \left\{ {\bm \Gamma}_\beta (E,V_\beta)  {\bf G} {\bm \Gamma}_{\alpha}(E,V_\alpha)  {\bf G}^\dagger  \right\}$ are integrated over the energy window defined by the difference of the Fermi functions \mbox{$f_\alpha(E)=\{ 1 + \exp[(E-E_F-eV_\alpha)/k_BT] \}^{-1}$} of macroscopic reservoirs into which semi-infinite ideal leads terminate. The matrices \mbox{${\bm \Gamma}_\alpha(E,V_\alpha)=i[{\bm \Sigma}_\alpha(E,V_\alpha) - {\bm \Sigma}_\alpha^\dagger(E,V_\alpha)]$} account for the level broadening due to the coupling to the leads [${\bm \Sigma}_\alpha(E,V_\alpha)$ are the self-energies introduced by the leads whose electronic structure is assumed to be rigidly shifted by the applied voltage $eV_\alpha$], thereby determining escape rates for electrons to leave the device and enter reservoirs where they are thermalized while the memory of their phase is lost.

Our MT-NEGF-DFT code utilizes  ultrasoft pseudopotentials and Perdew-Burke-Ernzerhof exchange-correlation functional. The localized basis set for DFT calculations  is constructed from atom-centered orbitals (six per C atom and four per H atom)  that are optimized variationally for the leads and the central molecule separately while their electronic structure is obtained concurrently.

{\em Zero source-drain bias regime.}---The interference effects on quantum transport in traditional metallic and semiconducting mesoscopic rings are typically studied by applying magnetic flux through the ring to change the quantum phase accumulated by an electron circulating around the ring~\cite{Nazarov2009}. However, this method is not applicable to explore quantum interference effects in molecular-size devices, since  magnetic field required to change the flux through the ring (e.g., from zero to $h/e$) far exceeds those available experimentally. Instead, one has to use particular geometry for attaching the electrodes to generate constructive or destructive quantum interference.  For example, in the absence of the third electrode in device setups of Fig.~\ref{fig:setup}, $\pi$-electron entering the molecule at the Fermi level has the wavelength $k_F/2d$ ($d$ is the spacing between carbon atoms within the molecule), so that for two simplest Feynman paths of length $10d$ (upper half of the ring) and $8d$ (lower half of the ring), the phase difference is $2k_Fd=\pi$ (the two Feynman paths are illustrated in the inset of Fig.~\ref{fig:setup}). To get the full transmission $T_{21}(E_F)$ between the source and drain electrodes requires a sum over all possible paths, including more complicated ones~\cite{Nazarov2009}, but they all cancel leading to a node $T(E_F)=0$. We note that this single particle explanation is insufficient to understand interference effects in very small ring-shaped molecules, such as benzene weakly coupled to gold electrodes, which are often dominated by Coulomb blockade effect so that degeneracies between many-body states of the isolated molecule have to be taken into account~\cite{Darau2009}.

The reference curve in Fig.~\ref{fig:transmission} is the zero-bias $T_{21}(E)$ for the two-terminal device, which is exponentially small at $E_F$, but still a {\em non-zero} quantity. Although exact transmission node $T_{21}(E_F) = 0$ is predicted for annulene-gold junctions using a phenomenological description~\cite{Cardamone2006}, the NEGF-DFT framework finds that tunneling of \mbox{$\sigma$-electrons} through hybridized \mbox{$\sigma$-orbitals} in the gold electrodes and \mbox{$\pi$-orbitals} in the molecule can wash out such transmission nodes for not too large molecules~\cite{Ke2008}. In devices presented here, the high contact transparency facilitates injection of evanescent states~\cite{Ke2007,Areshkin2009} from ZGNRs into the HOMO-LUMO energy gap of the isolated molecule. The overlap of such states in the middle of the molecule generates a transmission resonance near $E_F$, while destructive interference for (1,9) connection of ZGNR electrodes superimposes an antiresonance dip $T_{21}(E_F) \simeq 10^{-6}$ onto the two-terminal curve in Fig.~\ref{fig:transmission}.

When the infinite-impedance voltage probe 3  is attached to the upper half of 18-annulene, new incoherent Feynman paths emerge (such as $1 \rightarrow 3 \rightarrow 2$ illustrated in the inset of Fig.~\ref{fig:setup}) along which electrons propagate into the macroscopic reservoir through electrode 3 where they are dephased before entering the drain electrode 2. Such paths are not canceled, but instead generate~\cite{Nazarov2009} {\em incoherent} contribution $T_{21}^{\rm inc}(E)$ to the effective zero-bias transmission $\tilde{T}_{21}(E)$:
\begin{equation}\label{eq:trans}
\tilde{T}_{21}(E) = T_{21}(E) + T_{21}^{\rm inc}(E) = T_{21}(E) + \frac{T_{23}(E)T_{31}(E)}{T_{31}(E)+T_{23}(E)}.
\end{equation}
Figure~\ref{fig:transmission} shows that effective $\tilde{T}_{21}(E)$ increases by three orders of magnitude due to the attached voltage probe. Furthermore, when the third electrode is not bonded, but instead covers half of the ring as shown in Fig.~\ref{fig:setup}(b), $\tilde{T}_{21}(E) \simeq 0.4$ approaches the limit of a unit transmission through a single conducting channel (opened at $E_F$ of ZGNR electrodes), which has been one of the key goals in molecular electronics pursuits of a perfect link~\cite{Kiguchi2008}.

The role of the ``air-bridge" gate electrode in increasing the transmission by five orders of magnitude can be understood by examining the spatial profile of the local density of states (LDOS) in Fig.~\ref{fig:transmission} and charge density in Fig.~\ref{fig:charge}(b). They show how the top gate strongly modifies LDOS to enable charge transfer in the region underneath. These effects are due to the fact that the top gate is positioned at the distance $3.2$ \mbox{\AA} away from the planar device, so that hybridization due to overlap of gate orbitals with atomic orbitals underneath generates LDOS directly connecting leads 1 and 2.

{\em Finite source-drain bias regime.}---Driving mesoscopic and nanoscopic devices out of equilibrium by applying a finite bias voltage typically corrupts their quantum coherence (e.g., electrons at different energies within a bias voltage window have different interference patterns), so that standard interference effects are observed in the linear response regime~\cite{Nazarov2009}. Concurrently, at finite bias voltage $V_{ds}$ between the source and drain electrodes, the attached third electrode as in the device in Fig.~\ref{fig:setup}(a) would draw finite current~\cite{Cardamone2006}, which is an example of leakage currents whose minimization is one of the major tasks in downscaling FETs in digital electronics~\cite{Schwierz2010}. Here we use the setup in Fig.~\ref{fig:setup}(b) to investigate the range of modulation of source-drain current at \mbox{$V_{ds}=V_2-V_1=0.2$ V} via the top gate positioned at the distance 5.3 \mbox{\AA} away from the device. Figure~\ref{fig:finitebias} shows that current can be modulated from very small to finite value by changing the gate voltage $V_{gs}$, while the selected distance ensures that tunneling leakage current into the top electrode is very small. Thus, this type of a device could act as either a molecular-scale FET~\cite{Song2009} or as a bipolar junction transistor capable of amplifying current in the third lead.

{\em Conclusions.}---In conclusion, using NEGF-DFT formalism, recently extended to multiterminal devices~\cite{Saha2009a}, we analyzed quantum transport through ZGNR$|$18-annulene$|$ZGNR junction in the presence of the third ZGNR electrode attached either as the infinite-impendence voltage probe or serving as the ``air-bridge'' top gate covering half of the molecular ring. While both choices for the third electrode disrupt destructive quantum interferences (responsible for exponentially small transmission in the corresponding two-terminal device), transmission increases substantially further when using the ``air-bridge'' top gate which enhances the LDOS underneath to directly connect the source and the drain. Moreover, the latter device  in its `on' state  even approaches the long-sought limit~\cite{Kiguchi2008} in molecular electronics of a unit transmission through an organic-based device with a single conducting channel, because of high contact transparency brought by the proposed application of graphene-based electrodes.

\begin{acknowledgments}
We thank M. Grifoni and K. Richter for illuminating discussions. Financial support through NSF Grant No. ECCS 0725566 (K. K. S. and B. K. N.), DFG SFB 689 (during stay of B. K. N.  at the University of Regensburg) and DOE Grant No. DE-FG02-98ER45685 (W. L. and J. B.) is gratefully acknowledged. The supercomputing time for this research was provided in part by the NSF through TeraGrid resource TACC Ranger under Grant No. TG-DMR100002 (K. K. S. and B. K. N.).
\end{acknowledgments}




\begin{thebibliography}{10}

\bibitem{Wang2008a}
X.~R. Wang {\it et~al.}, Phys. Rev. Lett. {\bf 100},  206803  (2008).

\bibitem{Cai2010}
J. Cai {\it et~al.}, Nature {\bf 466}, 470 (2010).

\bibitem{Schwierz2010}
F. Schwierz, Nature Nanotech. {\bf 5}, 487 (2010).

\bibitem{Areshkin2007a}
D. Areshkin and C. White, Nano Lett. {\bf 7},  3253  (2007).

\bibitem{Kiguchi2008}
M. Kiguchi {\it et al.}, Phys. Rev. Lett. {\bf 101}, 046801 (2008); L. Venkataraman, Physics {\bf 1}, 5 (2008).

\bibitem{Cuniberti2005}
{\em Introducing Molecular Electronics}, edited by G. Cuniberti, G. Fagas, and K. Richter (Springer, Berlin, 2005).

\bibitem{Ke2007}
S.-H. Ke, H.~U. Baranger, and W. Yang, Phys. Rev. Lett. {\bf 99}, 146802  (2007).

\bibitem{Cervantes-Codi2008}
F. Cervantes-Sodi {\it et al.}, Phys. Rev. B {\bf 77}, 165427 (2008).

\bibitem{Song2009}
H. Song {\it et~al.}, Nature {\bf 462}, 1039 (2009).

\bibitem{Stokbro2003}
K. Stokbro {\it et~al.}, Comp. Mat. Science {\bf 27},  151 (2003).

\bibitem{Guo2006}
X. {Guo} {\it et~al.}, Science {\bf 311},  356  (2006).

\bibitem{Yazyev2008}
O.~V. Yazyev and M.~I. Katsnelson, Phys. Rev. Lett. {\bf 100},  047209 (2008).

\bibitem{Areshkin2009}
D.~A. Areshkin and B.~K. Nikoli\'{c}, Phys. Rev. B {\bf 79},  205430  (2009).

\bibitem{Jiang2008}
J. Jiang, W. Lu, and J. Bernholc, Phys. Rev. Lett. {\bf 101}, 246803 (2008).

\bibitem{Cardamone2006}
D. M. Cardamone, C. A. Stafford, and S. Mazumdar, Nano Lett. {\bf 6},  2422 (2006); C. A. Stafford, D. M. Cardamone, and S. Mazumdar, Nanotech. {\bf 18}, 424014 (2007).

\bibitem{Nazarov2009}
Y.~V. Nazarov and Y.~M. Blanter, {\em Quantum Transport: Introduction to Nanoscience} (Cambridge University Press, Cambridge, 2009).

\bibitem{Rincon2009}
J. Rinc\'{o}n {\it et al.}, Phys. Rev. Lett. {\bf 103},  266807  (2009); T. Hansen {\it et al.}, J. Chem. Phys. {\bf 131},  194704  (2009); T. Markussen, R. Stadler, and K. S. Thygesen, Nano Lett. {\bf 10}, 4260 (2010).

\bibitem{Areshkin2010}
D.~A. Areshkin and B.~K. Nikoli\'c, Phys. Rev. B {\bf 81},  155450  (2010).

\bibitem{Ke2008}
S.-H. Ke, W. Yang, and H.~U. Baranger, Nano Lett. {\bf 8},  3257  (2008).

\bibitem{Saha2009a}
K.~K. Saha {\it et al.}, J. Chem. Phys. {\bf 131},  164105  (2009).

\bibitem{Darau2009}
D. Darau {\it et al.}, Phys. Rev. B {\bf 79}, 235404  (2009).

\end{thebibliography}


\end{document}